\documentclass[aps,twocolumn,showpacs,preprintnumbers,nofootinbib,prl,superscriptaddress,groupedaddress,10pt]{revtex4-1}
\pdfoutput=1
\makeatletter
\def\l@subsubsection#1#2{}
\def\l@subsubsubsection#1#2{}
\makeatother

\usepackage{graphicx,amssymb,amsmath,amsthm,amsfonts,epsfig,epsf,fixmath}
\usepackage[usenames]{color}
\usepackage{epstopdf}
\usepackage{mathtools}
\usepackage{siunitx} 
\DeclareSIUnit \parsec {pc} 
\usepackage{aas_macros}
\usepackage{bm}
\usepackage{dcolumn}
\usepackage{latexsym}
\usepackage{rotating}
\usepackage{longtable}

\usepackage{enumerate}
\usepackage{tensor,multirow}
\usepackage{url}
\usepackage[linktocpage]{hyperref}

\newcommand{\totder}[2]{\frac{\mathrm{d} #1}{\mathrm{d} #2}}
\newcommand{\modulo}[1]{\left|#1\right|}

\begin{document}
\title{Model independent tests of the Kerr bound with extreme mass ratio inspirals}
\author{
Gabriel Andres Piovano$^1$,
Andrea Maselli$^1$,
Paolo Pani$^{1}$}

\affiliation{$^{1}$ Dipartimento di Fisica, ``Sapienza" Università di Roma \& Sezione INFN Roma1, Piazzale Aldo Moro 5, 
00185, Roma, Italy}


\begin{abstract} 
An outstanding prediction of general relativity is the fact that the angular momentum $S$ 
of an isolated black hole with mass $\mu$ is limited by the Kerr bound, $S\leq G\mu^2/c$.
Testing this cornerstone is challenging due to the difficulty in modelling spinning compact 
objects that violate this bound. We argue that precise, model-independent tests can be achieved by measuring 
gravitational waves from an extreme mass ratio inspiral around a supermassive object, 
one of the main targets of the future LISA mission.
In the extreme mass ratio limit, the dynamics of the small compact object depends only on its multipole moments, which 
are free parameters. At variance with the comparable-mass case, accurate waveforms are valid also when the 
spin of the small object greatly exceeds the Kerr bound. By computing the orbital 
dephasing and the gravitational-wave signal emitted by a spinning point particle in circular, nonprecessing, equatorial 
motion around a Kerr black hole, we estimate that LISA will be able to measure the spin of the small compact object at 
the level of $10\%$. Together with mass measurements, this will allow for theory-agnostic, unprecedented 
constraints on string-theory inspired objects such as ``superspinars'', almost in their entire parameter space.
\end{abstract}

\maketitle 

\noindent{{\bf{\em Introduction.}}}
The dawn of black-hole~(BH) physics can be arguably traced back to the seminal work 
by Penrose~\cite{Penrose:1962ij,Penrose:1964wq}, Wheeler~\cite{wheeler}, 
Hawking~\cite{Hawking:1976de}, Bekenstein~\cite{Bekenstein:1972tm,Bekenstein:1973ur}, 
Carter~\cite{Carter:1968ks,Carter:1968rr}, and many others during the first ``golden age'' 
of general relativity~(GR) in the 1970s. Since then, the field has evolved dramatically and 
several theoretical predictions have been experimentally confirmed to exquisite 
precision~\cite{Will:2014kxa,Yunes:2016jcc,TheLIGOScientific:2016src}. 
Nonetheless, some basic relativistic effects associated with BHs remain elusive and have 
not been yet tested directly. 

Arguably the most striking one is the fact that, within GR, a BH with mass $\mu$ can spin only 
below a critical value\footnote{The normalized 
bound, $S_{\rm max}/\mu^2=1$ (in $G=c=1$ units henceforth adopted~\cite{conventions}), 
is very modest: a solid ball of mass $1\,{\rm kg}$ and radius $10\,{\rm cm}$ making 
one revolution per second has $S/\mu^2\sim10^{17}$. On the other hand, a millisecond pulsar has $S/\mu^2\approx 
0.3$ (assuming standard values $\mu\approx 1.4M_\odot$ and moment of inertia $I\approx 10^{45}\,{\rm g\,cm}^2$).} of 
the angular momentum $S$, 
\begin{equation}
 |S|\leq S_{\rm max}\equiv \frac{G\mu^2}{c}\approx 
2\times 10^{44}\left(\frac{\mu}{15M_\odot}\right)^2\,{\rm kg}\,{\rm m^2/s}\,, \label{Kerrbound}
\end{equation}
above which a naked singularity would appear. Indeed, the unique stationary solution to GR in vacuum is 
the Kerr metric, which is regular outside an event horizon only if the above ``Kerr bound'' is fulfilled.
Therefore, any evidence of $|S|>S_{\rm max}$ in a compact object would imply either the presence of matter fields 
(e.g., compact stars can theoretical exceed the Kerr bound) or a departure from GR.

High-energy modifications to GR such as string theories can resolve curvature singularities, making the Kerr bound 
superfluous. Indeed, in these theories compact objects violating the bound~\eqref{Kerrbound} 
--~so-called \emph{superspinars}~-- arise generically~\cite{Gimon:2007ur}. A representative example is the large 
class of regular microstate geometries in supergravity theories 
(e.g.~\cite{Mathur:2008nj,Giusto:2012yz,Bena:2011fc,Bena:2015bea,Bianchi:2018kzy}).
These solutions have the same asymptotic metric of a Kerr BH, and their deviations in the near-horizon region are 
suppressed by powers of $M_P/\mu\ll1$, where $M_P$ is the Planck mass.
Therefore, besides the possible violation of the Kerr bound and its consequences (e.g. for the accretion efficiency of 
compact objects~\cite{Gimon:2007ur}), these solutions are practically indistinguishable from a BH (see 
Ref.~\cite{Cardoso:2019rvt} for a review).
In this context, testing the bound~\eqref{Kerrbound} provides a \emph{model-independent} way to test GR and 
high-energy extensions thereof.

However, testing the Kerr bound is very
challenging~\cite{Bambi:2011mj,Berti:2015itd,Yagi:2016jml,Barack:2018yly,Cardoso:2019rvt}. 
The standard route is to interpret observations in various contexts \emph{assuming} the Kerr metric and look for 
inconsistencies in explaining 
the data. This strategy is not optimal as one would wish to compare the Kerr case with 
some alternative and perform Bayesian model selection. The latter option is however 
hampered by the fact that the geometry of spinning BHs beyond GR~\cite{Berti:2015itd} --~or of spinning 
extreme compact objects without a horizon~\cite{Cardoso:2019rvt} 
(such as boson stars)~-- is known only perturbatively or 
numerically~\cite{Ryan:1996nk,Pani:2011gy,Kleihaus:2011tg,Herdeiro:2014goa,Ayzenberg:2014aka,Maselli:2015tta,
Barausse:2015frm,
Herdeiro:2016tmi,Cunha:2019dwb,Maselli:2015yva}. 
Furthermore, regardless of the technical difficulties, any analysis based on a specific model or theory 
would be limited to that specific case, whereas performing a model-independent test of 
the Kerr bound~\eqref{Kerrbound} would be much more profitable. In this respect, it is challenging to devise a test 
which is at the same time robust and sufficiently general.
For example, one could try to measure the spins of the
components of a comparable-mass binary using inspiral-merger-ringdown templates~\cite{LIGOScientific:2018mvr}; however, 
the latter \emph{assume} the two objects are Kerr BHs. 
Likewise, a generic post-Newtonian~\cite{Blanchet:2006zz} 
parametrization of the inspiral suffers 
from the fact that higher-order spin corrections enter at high post-Newtonian order, making the waveform of 
highly-spinning binaries less precise 
near coalescence. Furthermore, the post-Newtonian gravitational-wave~(GW) 
phase contains terms that explicitly assume the validity of the Kerr bound (e.g. the tidal heating 
term~\cite{Hartle:1973zz,PhysRevD.64.064004,Maselli:2017cmm}), so it is impossible to include violations consistently. 

In this letter and in a companion technical paper~\cite{companion}, we show that many of the above issues can be 
resolved with tests based on extreme mass ratio inspirals~(EMRIs), which are also model independent to a large extend.
EMRIs are one of the main targets of the future space-based 
Laser~Interferometer~Space~Antenna~(LISA)~\cite{Audley:2017drz} and of evolved 
concepts thereof~\cite{Baibhav:2019rsa}. 
Owing to the large number of GW cycles, EMRI signals detectable by LISA can be used 
to extract the binary parameters with exquisite accuracy~\cite{Babak:2017tow},
and to perform unique tests of fundamental 
physics~\cite{Glampedakis:2005cf,Barack:2006pq,Sopuerta:2009iy,Yunes:2011aa,Pani:2011xj,Barausse:2016eii,
Chamberlain:2017fjl,Babak:2017tow,Cardoso:2018zhm,Pani:2019cyc, Datta:2019epe} (see 
\cite{Barack:2018yly,Barausse:2020rsu} for some recent reviews).

Remarkably, EMRIs allow to devise tests which do not require any assumption on the 
specific properties of the secondary other than specifying its multipole moments. This is a great advantage to study 
generic (and arguably vague) proposals such as the superspinar one.

\noindent{{\bf{\em Setup.}}}
In an EMRI a small, stellar-size, compact object (dubbed as secondary) of 
mass $\mu$ orbits around a supermassive object (dubbed as primary) of mass 
$M \sim(10^6 - 10^9) M_\odot$; the typical mass ratio of the system is 
$q = \mu/M\in (10^{-7} - 10^{-4})$ and therefore allows for a small-$q$ 
expansion of Einstein's equations. 
To the leading order in $q$, the dynamics is described by a point particle 
of mass $\mu$ in motion around the primary. The orbits evolve quasi-adiabatically 
through a sequence of geodesics due to energy and angular momentum loss carried 
away by GWs~\cite{Poisson:2003nc,Barack:2009ux}. 
To higher order in a small-$q$ expansion, the dynamics can still be described by a point particle 
endowed with a series of multipole moments. The post-adiabatic corrections depend on 
self-force effects~\cite{Barack:2009ux} and also on the intrinsic angular momentum $S$ of the secondary, 
which is the main target of our investigation. It is convenient to introduce the dimensionless quantity
\begin{equation}
\sigma \equiv \frac{S}{\mu M} = \chi q\ ,\label{def:sigma}
\end{equation}
where $\chi \equiv S/\mu^2$ is the reduced spin of the secondary. 
Owing to the mass ratio 
dependence, for an EMRI $|\sigma|\ll 1$ even when $\chi$ is very large, since it is sufficient that 
$|\chi|\ll1/q \sim (10^4-10^7)$. Therefore, it is possible to linearize the dynamics to ${\cal O}(\sigma)$ even 
when $\chi$ is large. This is an enormous advantage relative to other cases, e.g. 
post-Newtonian waveforms of comparable mass binaries.
Thus, in order to test the Kerr bound~\eqref{Kerrbound} we can study the EMRI evolution in which 
the secondary is assumed to be either a Kerr BH, which fulfills the constraint  
$|\chi|\leq1$, or an extreme compact object~\cite{Cardoso:2019rvt}
that can violate such a bound, i.e. $\modulo{\chi}>1$.

High-curvature corrections for the primary are negligible compared to the secondary. For concreteness, let us consider 
an effective field theory described by the Einstein-Hilbert Lagrangian with higher-order curvature 
terms~\cite{Berti:2015itd} in the schematic form
\begin{equation}
 {\cal L} = R+ \beta {\cal R}^n +.... \,, \label{Lagrangian}
\end{equation}
where $n>1$, $R$ is the Ricci scalar, ${\cal R}$ is some curvature scalar operator, and $\beta$ is a coupling constant 
with dimensions of $({\rm mass})^{2n-2}$. It is easy to see that the corrections to the primary are suppressed by a 
factor $1/q^n\gg1$ or higher~\cite{companion}. Therefore, to an 
excellent approximation we can assume that the background spacetime (i.e. the primary) is described by the Kerr metric 
with mass $M$ and angular momentum $J\equiv \hat a M^2$ satisfying the Kerr bound, i.e. 
$\vert \hat a\vert \leq 1$.

The dynamics of the spinning point particle on the Kerr background can be obtained through the covariant conservation 
of the energy-momentum tensor leading to the Mathisson-Papapetrou-Dixon equations of 
motion~\cite{Mathisson:1937zz,Papapetrou:1951pa, 
Corinaldesi:1951pb,Tulczyjew:1959,Dixon:1964NCim,Dixon:1970I,Dixon:1970II,Steinhoff:2009tk}. 
We integrate such equations supplied by the Tulczyjew-Dixon condition, $S^{\mu\nu}p_\nu=0$, 
between the spin tensor $S^{\mu\nu}$ and the body 4-momentum $p^\mu$~\cite{Kyrian:2007zz}. This 
constraint fixes the center of mass reference frame, and guarantees that 
the squared mass $\mu^2=-p_\mu p^\mu$ and spin magnitude $S^2=\frac{1}{2}S_{\mu\nu}S^{\mu\nu}$ are conserved during 
the orbital evolution.

During its motion the secondary acts as a perturbation of the background spacetime. Within GR, GW emission from the 
binary can be computed using the Teukolsky formalism~\cite{Teukolsky:1972my,Drasco:2005kz,Hughes:1999bq}.
\begin{equation}
\Delta^2\frac{d}{dr}\left(\frac{1}{\Delta}\frac{d R_{\ell m\omega}}{dr}\right)-V_{\ell m\omega}(r)R_{\ell m 
\omega}={\cal T}_{\ell 
m\omega}(r)\,,\label{waveeq}
\end{equation}
for any integer $\ell\geq2$ and $|m|\leq\ell$. The effective potential $V_{\ell m\omega}(r)$ is given in 
Ref.~\cite{Barack:2009ux}, whereas the source term ${\cal T}_{\ell m\omega}(r)$ depends on the stress-energy tensor of 
the secondary. The latter depends explicitly on the spin 
$\sigma$ in two ways: directly, since the spin of the secondary affects the energy content of the source, and 
indirectly through the trajectory of the secondary, which is affected by spin-angular momentum couplings. The final 
expression is cumbersome and derived in detail in Ref.~\cite{companion}.

We shall neglect extra radiation channels and consider only the standard GW emission in GR. The motivation for this 
choice is twofold: (i) superspinars can also arise within GR in the presence of exotic matter fields, in which case our 
analysis is exact; (ii) in case of  higher-curvature corrections to GR as in Eq.~\eqref{Lagrangian} the ordinary GW 
emission remains unchanged in the EMRI limit, but there might be further dissipation channels (e.g. dipolar radiation) 
in case the secondary is charged under a massless field~\cite{Maselli:2020zgv}. However, in the context of supergravity 
and string theories, putative extra degrees of freedom are expected to be extremely heavy and therefore do not 
propagate at the frequency of an EMRI. Thus, corrections to the dissipative sector are also negligible. At any 
rate, extra putative dissipative channels (e.g. due to massless degrees of freedom) can be straightforwardly 
accommodated within our framework.

\noindent{{\bf{\em GW flux.}}}
We numerically integrate Eq.~\eqref{waveeq} using a standard Green function approach~\cite{companion}, which
allows us to compute the energy fluxes down to horizon, $\dot{E}_\textnormal{GW}^{-}$, and at infinity, 
$\dot{E}_\textnormal{GW}^{+}$. The impact of the spin of the secondary on the GW fluxes has been studied in 
Refs.~\cite{Tanaka:1996ht,Burko:2003rv,Burko:2015sqa,Harms:2015ixa,Harms:2016ctx, 
Lukes-Gerakopoulos:2017vkj,Warburton:2017sxk,Akcay:2019bvk,Yunes:2010zj,Chen:2019hac}. 
A detailed analysis of the fluxes and a 
comparison with previous work~\cite{Harms:2015ixa,Akcay:2019bvk,Yunes:2010zj,Taracchini:2013wfa,Gralla:2015rpa} is presented in Ref.~\cite{companion}.

In the EMRI limit the radiation-reaction time scale is much 
longer than the orbital period. We can therefore assume that the inspiral is quasi-adiabatic. 
Under this approximation the system evolves as the change in the binding 
energy $\dot{E}_\textnormal{orb}$ is balanced by the total GW flux at infinity and at the 
horizon, 
\begin{equation}
 -\dot{E}_\textnormal{orb}=\dot{E}_\textnormal{GW}=\dot{E}_\textnormal{GW}^{+}+\dot{E}_\textnormal{GW}^{-}\,.
\end{equation}
%
The flux balance law is valid also when the spin of the secondary is taken into account~\cite{Akcay:2019bvk}.
Assuming that the secondary spin remains constant during the evolution, angular-momentum fluxes are directly 
related to the energy ones~\cite{Kennefick:1998ab,Tanaka:1996ht,companion}.
All fluxes are decomposed in multipole modes [$\ell=2,3,..$ in Eq.~\eqref{waveeq}].
In our calculations we sum the multipole contributions up to $\ell=20$; 
truncation errors are $0.05\%$ in the most extreme cases, i.e. $\hat a=0.995$ at the innermost stable circular 
orbit~(ISCO), and typically much smaller~\cite{companion}. 

Furthermore, in the adiabatic approximation a two-time scale analysis shows that during the EMRI evolution the masses 
and spins of the binary can be neglected to leading order~\cite{Hinderer:2008dm}. Likewise, the evolution of the spin 
of the secondary --~which introduce dissipative self-torque~\cite{Akcay:2019bvk}~-- is subdominant 
with respect to the effects discussed here~\cite{companion}. 

\noindent{{\bf{\em Results.}}}
%
\begin{figure}[t]
 \centering
 \includegraphics[width=0.45\textwidth]{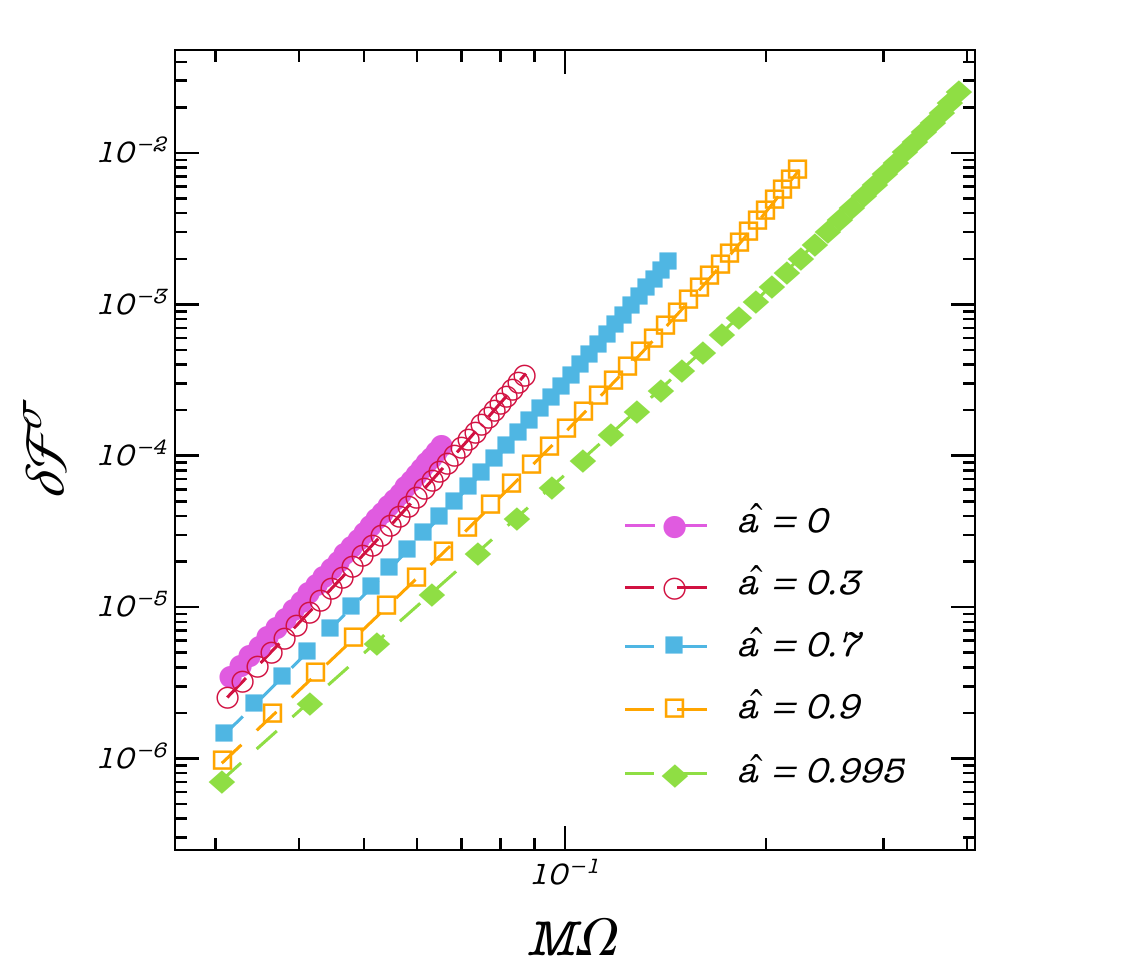}
 \caption{The spin-correction coefficient $\delta \dot E^\sigma_\text{GW}$ [see Eq.~\eqref{EdotGW}] as a function of 
the orbital frequency (up to the ISCO) for different values of the spin $\hat a\equiv J/M^2$ of the primary.}
 \label{fig:angfluxes} 
\end{figure}
In the $|\sigma|\ll1$ limit, the GW flux can be expanded at fixed orbital frequency $\Omega$ as
\begin{equation}
\dot E_{\text{GW}} = q^2\left({\dot E}^0_{\text{GW}} + \sigma \delta {\dot E}^\sigma_{\text{GW}} +{\cal 
O}(\sigma^2)\right)\,, \label{EdotGW}
\end{equation}
where we have factored out an overall mass-ratio dependence, ${\dot E}^0_{\text{GW}}$ is the (normalized) flux for a 
nonspinning secondary, whereas $\delta {\dot E}^\sigma_\text{GW}$ is 
the (normalized) spin contribution of the secondary. As 
anticipated, the latter is suppressed by a factor of ${\cal O}(q)$ relative to the leading-order term. It therefore 
enters at the same order as the leading-order conservative part and the second-order dissipative part of the 
self-force~\cite{Barack:2009ux,Pound:2012nt}. In Fig.~\ref{fig:angfluxes} we show $\delta {\dot 
E}^\sigma_\text{GW}$ as a function of the orbital frequency, $\Omega=\Omega(r)$. As expected, the correction becomes 
stronger when the orbit approaches the ISCO (i.e.  for higher frequencies) and when the primary is rapidly spinning.

\begin{figure}[t]
\centering
\includegraphics[width=0.43\textwidth]{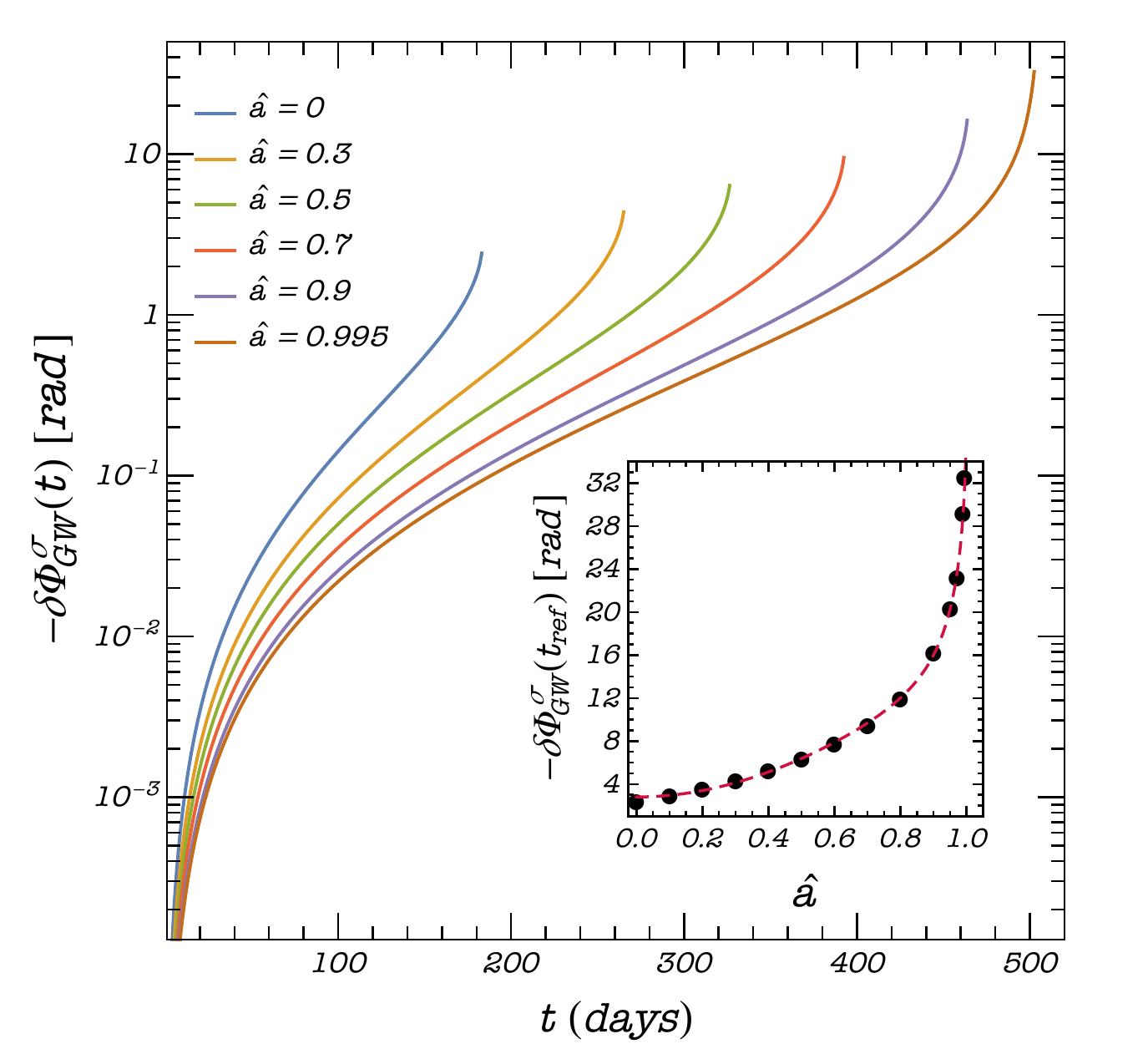}
\caption{Spin 
correction, $\delta \Phi^\sigma_{\rm GW}(t)$, to the 
instantaneous GW phase [cf. Eq.~\eqref{eq:instphase}] as a function of time up to the ISCO for different values of the 
spin $\hat a$ of the primary. The inset shows the spin correction, 
$\delta \Phi^\sigma_{\rm GW}(t_{\rm ref})$, to the total accumulated phase. The dashed colored curve shows the 
fit~\eqref{fit}. We assumed $\mu = 30 M_\odot$ and $M = 10^6 M_\odot$ as reference values. 
Note that in general $\delta \Phi^\sigma_{\rm GW}(t_{\rm ref})<0$, i.e. when $\chi>0$ the inspiral lasts longer.
Data are publicly available~\cite{webpage}.
}
\label{fig:phase} 
\end{figure}

With the fluxes $\dot E_{\text{GW}}$ at hand, it is possible to calculate the evolution of the orbital frequency 
${r}(t)$ and phase $\Phi(t)$ due to radiation losses. In the adiabatic approximation, 
\begin{alignat}{2}
\totder{r}{t} &= - \dot E_{\text{GW}}(r) \bigg( \totder{{E}}{{r}}\bigg)^{\!-1}\,, \qquad 
\totder{\Phi}{t} &= \Omega(r)\,, \label{eqdephasing}
\end{alignat}
where the particle energy $E$ and the angular velocity $\Omega$ are analytical (albeit cumbersome) functions 
of ($r, \hat a, \sigma$)~\cite{Steinhoff:2012rw,Jefremov:2015gza,companion}.
The flux $\dot E_{\text{GW}}(r)$ is obtained by interpolating the calculated fluxes in the range 
$r\in(r_\text{start},r_\text{ISCO})$.
The starting point $r_\textup{start}$ is chosen such that the initial orbital frequency  is the same as in the case of a 
nonspinning particle at the reference value $r = 10.1 M$.

By integrating the system~\eqref{eqdephasing} we can obtain the \emph{instantaneous} orbital phase, which is related 
to the GW phase of the dominant mode by $\Phi_{\rm GW}=2\Phi$. The latter can be schematically written as 
\begin{equation}
\Phi_{\rm GW} (t) = \frac{1}{q}\left(\Phi^0_{\rm GW}(t) +  \sigma \delta \Phi^\sigma_{\rm GW}(t) +{\cal 
O}(\sigma^2)\right)\,, \label{eq:instphase}
\end{equation}
where $\Phi_{\rm GW}^0(t)$ is the (normalized) phase for a nonspinning secondary, $\delta \Phi_{\rm 
GW}^\sigma(t)$ is the (normalized) spin correction, and we have factored out an overall $q^{-1}$ dependence.
As expected, the spin correction is suppressed by a factor of ${\cal 
O}(q)$ and is therefore independent of $q$ to the leading order, since the factor $q^{-1}$ cancels out with $\sigma$ 
[see Eq.~\eqref{def:sigma}]. In Fig.~\ref{fig:phase} we show $\delta \Phi^\sigma_{\rm GW}(t)$ and the 
spin correction to the \emph{accumulated} GW phase, $\Phi^{\rm tot}_{\rm GW} =\Phi_{\rm GW}(t_{\rm ref})$, for some 
representative examples. Our results are in overall agreement with previous analyses that used ``kludge'' or 
effective-one-body 
waveforms~\cite{Barack:2006pq,Huerta:2011kt,Yunes:2010zj}, which however rely to some extend on the post-Newtonian 
approximation and might fail to describe accurately the contribution of small effects in the late-time EMRI dynamics.
As a reference, we chose the same $t_{\rm ref}$ for any value of $\chi$. In particular, we chose
$t_{\rm ref}$ as the time to reach the ISCO for a nonspinning secondary minus $0.5\,{\rm day}$, so that the evolution 
stops before the transition from inspiral to plunge (occurring near the ISCO~\cite{Ori:2000zn,Burke:2019yek}) for any 
value of $\chi$ and $\hat a$. A useful fit of the total accumulated GW 
phase is: 
\begin{equation}
 \delta \Phi^\sigma_{\rm GW}(t_{\rm ref}) = \sum^3_{i=0} c_i (1-\hat a^2)^{i/2} + c_4\hat a\ ,\label{fit}
\end{equation}
where $c_0=38.44, c_1=-90.36, c_2=99.43, c_3=-44.95, c_4=1.91$. The fit is accurate within 
$5\%$ in the whole range $\hat a\in[0,0.995]$, with better accuracy at large $\hat a$.

\noindent{{\bf{\em Measuring the spin of the secondary.}}}
Parameter estimation for EMRIs is a challenging problem~\cite{Babak:2017tow,Chua:2019wgs}, especially if one wishes to 
use exact numerical waveforms rather than approximate ones. Here we estimate 
the potential to measure $\chi$ by using a simple requirement: a total dephasing $\approx 
1\,{\rm rad}$ or greater is likely to substantially impact a matched-filter search, leading to a significant loss 
of detected events~\cite{Lindblom:2008cm}. A more rigorous parameter-estimation analysis for the spin of the secondary 
is in progress and will appear elsewhere.

Let us consider two waveforms which differ only by the value of the spin of the secondary, $\chi_A$ 
and $\chi_B$, respectively. Using Eq.~\eqref{eq:instphase}, the minimum difference $\Delta \chi=\chi_B-\chi_A$ 
which would lead to a difference in phase at least of $\alpha\,{\rm rad}$ is:
\begin{equation}
 |\Delta\chi|> \frac{\alpha}{|\delta\Phi^\sigma_{\rm GW}|}\ .\label{criterion}
\end{equation}
For a reference value $\hat a=0.7$ 
($\hat a=0.9$) with $\alpha=1$~\cite{Lindblom:2008cm}, we obtain $|\Delta \chi|>0.1$ ($|\Delta 
\chi|>0.05$). 
Thus, our simplified analysis shows that EMRIs can provide a measurement of the spin of the secondary at the 
level of $5-10\%$ for fast spinning primaries. This adds to the outstanding accuracy in the measurements 
of $M$, $\hat a$, and $\mu$~\cite{Barack:2006pq,Babak:2017tow}.
More stringent constraints would arise by an analysis of the mismatch between two 
waveforms~\cite{Flanagan:1997kp,Lindblom:2008cm}. Requiring the latter to be smaller than $\sim 
1/(2\rho^2)$, where $\rho$ is the signal-to-noise ratio of the event, suggests using $\alpha<1$ for 
back-of-the-envelope estimates~\cite{Datta:2019epe}. 
\begin{figure}[th]
\centering
\includegraphics[width=0.49\textwidth]{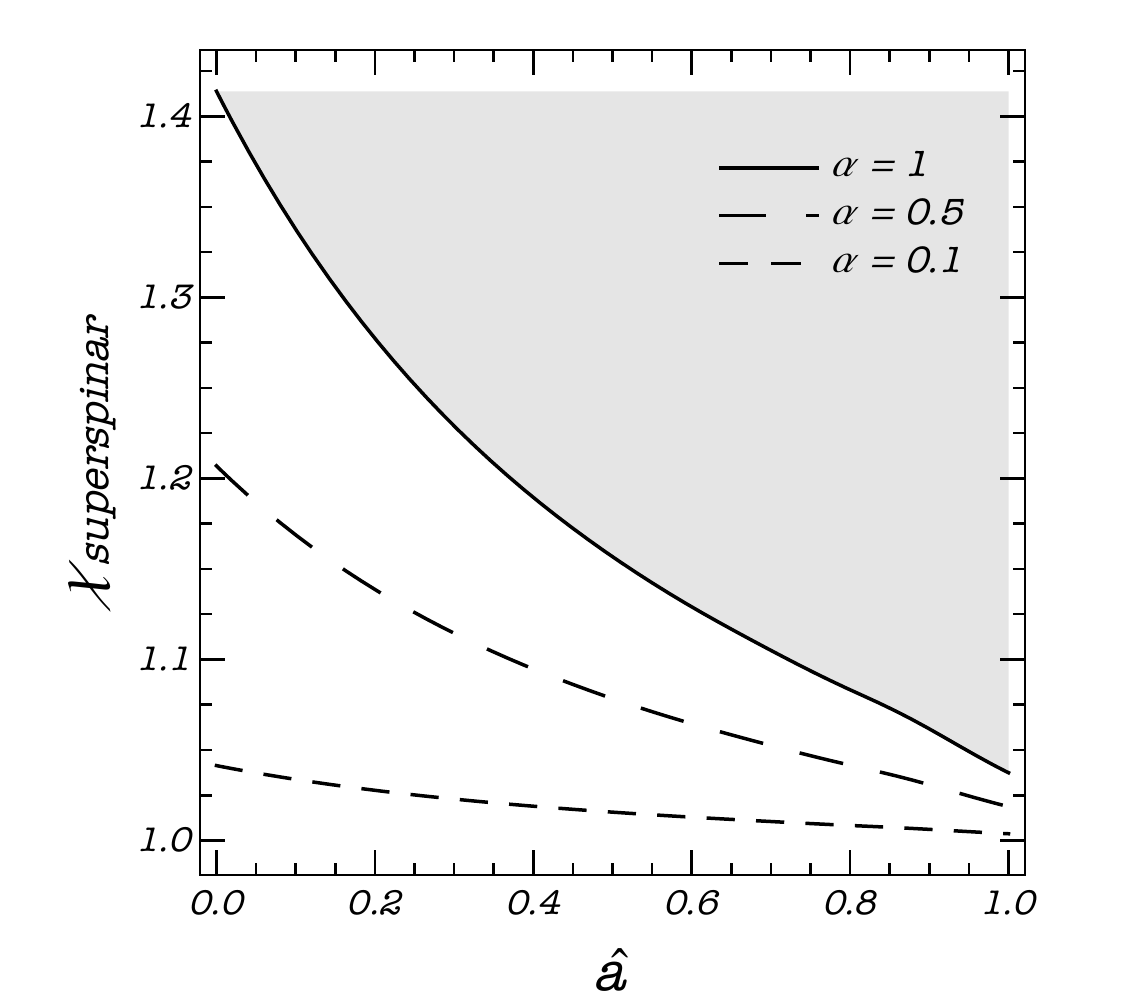} 
\caption{Exclusion plot  for the spin of a superspinar obtained using the criterion~\eqref{criterion}. A 
measured dephasing at the level of $\alpha\,{\rm rad}$ would exclude/probe the region above each curve.}
\label{fig:exclusion} 
\end{figure}

\noindent{{\bf{\em Superspinars as hints of new physics.}}}
In addition to providing an accurate and model-independent way to measure spins of stellar-mass 
objects~\cite{Barack:2006pq,Huerta:2011kt,Yunes:2010zj,companion}, EMRI detections can provide theory-agnostic tests of 
the Kerr bound~\eqref{Kerrbound} and, in particular, of superspinars~\cite{Gimon:2007ur}. 

As previously discussed, $\Delta\chi<1$ for any value of $\hat a$, suggesting that 
the spin of a fast spinning Kerr secondary could be measured with an accuracy better than $100\%$. For 
reference values $\chi\approx\hat a\approx 0.7$, the accuracy is approximately $15\%$, which would exclude $\chi>1$ 
at $3\sigma$ confidence level.
Indeed, since $|\chi_A|\leq 1$ for a Kerr BH, an accuracy at the level of 
(say) $|\Delta \chi|>0.1$ allows us to distinguish a Kerr BH from another fast spinning object provided the spin of 
the latter is $|\chi_B|\gtrsim 1+|\Delta\chi|\approx 1.1$. In Fig.~\ref{fig:exclusion} we show the exclusion plot for 
the spin of a superspinar obtained using the criterion~\eqref{criterion} and under the most conservative assumption, 
$\chi_A=1$, as a function of the spin $\hat a$ of the primary. We consider different values for the dephasing threshold 
$\alpha$. For the standard choice of $\alpha =1\,{\rm rad}$, our results suggest that it should be possible to 
exclude/probe the range $|\chi_B|>1.4$ ($|\chi_B|>1.05$) for nonspinning (highly-spinning) primaries.  
Since no theoretical upper bound is expected for superspinars (other, possibly, than those coming from the 
ergoregion instability~\cite{Pani:2010jz,Maggio:2017ivp,Maggio:2018ivz,Roy:2019uuy}) a spin measurement at this level 
can potentially probe a vast region of the parameter space for these objects.

One might argue that, while clearly incompatible with the secondary 
being a Kerr BH, a putative EMRI measurement of $\chi>1$ could still be compatible with the secondary being a neutron 
star or a white dwarf. 
Given that neutron stars and white dwarfs have masses in the narrow range $\mu\sim (1-2)M_\odot$, an EMRI measurement 
of $\mu$ --~surely available for events favorable enough to measure $\chi$~-- in the range $\mu>3M_\odot$ or $\mu\ll 
M_\odot$ would exclude a standard origin for the superspinar. Furthermore, even in the case in which 
$\mu\in(1,2)M_\odot$, the spin of an isolated compact star is expected to be significantly smaller than 
the Kerr bound. As a reference, the spin of the fastest pulsar known to date is
$\chi\approx0.3$~\cite{Hessels:2006ze}. Out of $340$ observations of millisecond pulsars in the ATNF Pulsar 
Database~\cite{Manchester:2004bp}, $\langle \chi\rangle = 0.11\pm0.04$, suggesting that $|\chi|>1$ would be very 
unlikely\footnote{It is also interesting to note 
that there is no solid explaination for the fact that the angular velocity of observed 
pulsars is systematically well below the theoretical (mass-sheedding) limit. In this context EMRIs can provide a 
model-independent portal to discover neutron stars spinning faster than the current population,} . Isolated white dwarfs 
have comparable values of $\chi$. The fastest spinning white dwarf to date has 
$\chi\approx 10$, but it is strongly accreting from a binary companion~\cite{1997A&A...317..815B}.
Less compact objects (such as brown dwarfs) could spin faster then the Kerr bound, but are tidally disrupted 
much before reaching the ISCO~\cite{companion} so they can also be easily distinguishable from exotic superspinars.

\noindent{{\bf{\em Discussion.}}}
EMRIs are unique probes of fundamental physics~\cite{Barack:2018yly,Barausse:2020rsu}. Besides 
offering the opportunity for exquisite tests of 
gravity~\cite{Sopuerta:2009iy,Yunes:2011aa,Pani:2011xj,Barausse:2016eii,Chamberlain:2017fjl,Cardoso:2018zhm} and of the 
nature of supermassive 
objects~\cite{Barack:2006pq,Pani:2010em,Babak:2017tow,Pani:2019cyc,Datta:2019epe}, here we have 
shown that they can be used to perform theory-agnostic tests of the Kerr bound. Our results suggest that EMRI 
detections with LISA have the potential to rule out (or detect) superspinars almost in the entire region of the 
parameter space.
This conclusion is based on a simplistic analysis, which must be validated with a more 
careful study, for example including accurate waveform models, a statistical analysis that can 
account for correlations among the waveform parameters~\cite{Huerta:2011kt}, and the fact that LISA will be a 
signal-driven GW detector, so that numerous simultaneously-detected sources must be suitably 
subtracted~\cite{Audley:2017drz,Chua:2019wgs,LISADataChallenge}. 

Given the fact  that the secondary spin is a small effect, a faithful measurement requires having all first-order 
post-adiabatic effects under control. At the same time, no EMRI inspiral and post-adiabatic waveform model is complete 
without including the spin of the secondary along with first-order conservative and second-order dissipative self-force 
effects~\cite{Akcay:2019bvk}. In addition to a more rigorous parameter estimation including also self-force effects 
and possible confusion with environmental effects (although the latter are typically 
negligible~\cite{Barausse:2014tra}), future work will focus on noncircular/nonequatorial orbits and on the case of 
misaligned spins, which introduces precession in the motion~\cite{Tanaka:1996ht}.

Finally, it would be very interesting to include higher multipole moments for the secondary, in particular the 
quadrupole moment. Although this effect is suppressed by a further ${\cal O}(q)$ factor and is probably too small to 
be detectable with LISA, it can potentially allow performing model-independent tests of the no-hair theorem on the 
secondary.

\noindent{{\bf{\em Acknowledgments.}}}
We thank Richard Brito and Niels Warburton for useful discussion.
This work makes use of the Black Hole Perturbation Toolkit and \textsc{xAct} \textsc{Mathematica} package. 
P.P. acknowledges financial support provided under the European Union's H2020 ERC, Starting 
Grant agreement no.~DarkGRA--757480, and under the MIUR PRIN and FARE programmes (GW-NEXT, CUP:~B84I20000100001).
The authors would like to acknowledge networking support by the COST 
Action CA16104 and support from the Amaldi Research Center funded by 
the MIUR program "Dipartimento di Eccellenza" (CUP: B81I18001170001).
%

\bibliographystyle{utphys}
\bibliography{../Ref}

\begin{thebibliography}{100}
\bibitem{conventions}
We use geometric units, $G=c=1$, and we define the Riemann tensor as
  ${R_{\mu\nu\sigma}}^\delta \omega_\delta =2
  \nabla_{[\mu}\nabla_{\nu]}\omega_\sigma$, where $ \nabla_\mu$ is the
  covariant derivative and $\omega_\delta$ an arbitrary 1-form, while the
  square brackets denotes the antisymmetrization. This is the same notation
  adopted in the package \textsc{xCoba} of the software \textsc{Mathematica},
  which we used for all tensor computation. The metric signature is
  $(-,+,+,+)$.
\end{thebibliography}

\end{document}